\documentclass[singlespacing]{elsart}

\usepackage{graphicx}

\usepackage{amssymb}
\journal{}
\begin{document}

\begin{frontmatter}

\title{Low frequency limit for thermally activated
 escape with periodic driving}

\author{A.E. Sitnitsky},
\ead{sitnitsky@mail.knc.ru}

\address{Institute of Biochemistry and Biophysics, P.O.B. 30, Kazan
420111, Russia}

\begin{abstract}
The period-average rate in the low frequency limit for thermally activated
 escape with periodic driving is derived
in a closed analytical form. We define the low frequency limit as the one
where there is no essential dependence on frequency so that the formal limit
$\Omega \rightarrow 0$ in the appropriate equations can be taken.
 We develop a perturbation theory of
the action in the modulation amplitude and obtain a cumbersom but
closed and tractable formula for arbitrary values of
the modulation ampitude to noise intensity ratio $A/D$
except a narrow region near the bifurcation
point and a
 simple analytical formula for the limiting case of
 moderately strong modulation. The present theory yields
 analytical description for the retardation of the exponential
growth of the escape rate enhancement (i.e., transition from a log-linear
regime to more moderate growth and even reverse behavior).
The theory is developed for an arbitrary potential with an activation
barrier but is exemplified by the cases of cubic (metastable) and quartic
(bistable) potentials.

Keywords: Kramers' theory, thermally activated escape,
 periodic driving.
\\

\end{abstract}

\end{frontmatter}

\section{Introduction}
Thermally activated escape over a potential barrier is ubiquitous
in physics, chemistry and biology
 (see \cite{Han90}, \cite{Jun93}, \cite{Fle93}, \cite{Tal95},
 \cite{Gam98} and refs. therein). This phenomenon is important for
both quantum \cite{Han90}, \cite{Jun93}, \cite{Wei99}, \cite{Ank01},
\cite{Ank01a}, \cite{Ank05} and classical \cite{Han90},
\cite{Jun93}, \cite{Rei97}, \cite{Leh00},
  \cite{Leh000}, \cite{Leh03}, \cite{Sme99},
\cite{Sme99a}, \cite{Mai01},\cite{Dyk01}, \cite{Tal99}, \cite{Tal04},
\cite{Ryv04}, \cite{Dyk04},
 \cite{Dyk05}, \cite{Dyk051} systems. It can proceed in strong friction
(overdamped), weak friction (underdamped) and needless to say intermediate
 regimes (see refs. above). The case of thermally activated escape
unperturbed by additional external influences pioneered by Kramers
is exhaustively investigated and by now is a well
understood phenomenon.

However in most physical realizations the thermally activated escape is
 modulated by some external driving. In this case the  stationary limit
that may be a rather good approximation in the absence of external
driving is already inapplicable. In the presence of the latter the system becomes
intrinsically non-equilibrium and the problem in known to be
notoriously difficult for analytical treatment \cite{Jun93}.
 The particular
 case of periodically driven escape \cite{Jun93}, \cite{Rei97},
 \cite{Leh00},
  \cite{Leh000}, \cite{Leh03}, \cite{Sme99},
\cite{Sme99a}, \cite{Mai01}, \cite{Dyk01}, \cite{Tal99}, \cite{Tal04},
\cite{Ryv04}, \cite{Dyk04},
 \cite{Dyk05}, \cite{Dyk051} is relevant among others for
 chemical physics \cite{Gho05}, \cite{RdV01}, \cite{Sha00}
 (where a chemical reaction can be influenced by, e.g., laser electric field)
 and enzymology \cite{Sit06} (where an enzymatic reaction can be influenced by
an oscillating electric field produced by the dynamics of protein
 structure \cite{Sit07}).
 According to
\cite{Dyk01} the periodic driving force ""heats up" the system by changing
its effective temperature thus giving rise to lowering of the activation energy
of escape which can be much bigger than the real temperature even for comparatively
 weak fields".

 Revealing the
physical aspects of enzyme action may well become one of the most
important grounds for application of the periodically modulated thermally
activated escape theory. In support of this point of view it is worthy to
note that understanding the role of driving at activated escape in biological
systems is considered by the authors of \cite{Dyk01} as "a fundamentally
important and most challenging open scientific problem". The reasons for the
above statement are as follows. The problem of enzyme catalysis is the main
unsolved interdisciplinary enigma of molecular biophysics,
biochemistry and needless to say enzymology. Up to now there
is no definite and commonly accepted understanding of "how does an
enzyme work?" (see, e.g., heat controversy at a recent conference in
the subject issue of Phil. Trans. R. Soc. B (2006) 361). The idea
that dynamical effects may play a crucial role at enzyme action is
very popular at present and the concept of the so-called rate
 promoting
vibration is a central one in modern enzymology. However the
 practitioners engaged in
chemical enzymology traditionally discuss the phenomenon of
enzyme catalysis in terms and notions of the transition state
theory (as can be seen
from the materials of the above mentioned conference). The latter
 is essentially
equilibrium one that is embedded into its cornestone postulate
and is poorly suited for taking into account dynamical effects in
a reaction rate.
The main tool to study such effects is the Kramers' theory
 \cite{Han90}, \cite{Jun93}, \cite{Gam98}, \cite{Ray99}. Regretfully
at present this theory is much less known and necessitated for applications
in enzymology than its transition state theory counterpart.

The thermally activated escape problem at periodic driving in the
overdamped classical regime was conceptually solved in the papers
\cite{Leh00}, \cite{Leh000}, \cite{Leh03}, \cite{Sme99},
\cite{Sme99a}, \cite{Mai01}, \cite{Tal99}, \cite{Tal04},
\cite{Ryv04}, \cite{Dyk04}, \cite{Dyk05}, \cite{Dyk051}. Two
mutually complementary theories \cite{Leh00}, \cite{Leh000},
\cite{Leh03} and \cite{Sme99}, \cite{Sme99a}, \cite{Ryv04},
\cite{Dyk04}, \cite{Dyk05}, \cite{Dyk051} provide deep
insights on the behavior of many physical values of
interest. They are based on a physical idea of the optimal
 path and are
able to provide an imaginable picture of the process.
The papers \cite{Jun93}, \cite{Sme99}, \cite{Dyk01}
 provide description of the
escape rate enhancement at weak modulation (the so-called log-linear
regime) where the change of the activation energy is linear in the
 modulation amplitude.
Most important of all, the escape rate enhancement
 exhibits the replacement of the log-linear regime by more
 moderate growth with the increase of modulation amplitude to noise
intensity ratio. Such behavior for the intermediate regime of moderately
strong and moderately fast driving
is well described by the \cite{Leh00},
 \cite{Leh000} theory that is corroborated by high-precision numerical
results. The scaling bahavior of
the prefactor near the bifurcation point is investigated
 in details \cite{Ryv04}, \cite{Dyk04}, \cite{Dyk05}, \cite{Dyk051}.
In particular the so called adiabatic regime is
 most thorougly investigated \cite{Jun93}, \cite{Tal99},
\cite{Dyk05}, \cite{Dyk051}. This regime is defined
by the authors of \cite{Sme99}, \cite{Sme99a}, \cite{Dyk01},
\cite{Ryv04}, \cite{Dyk04}, \cite{Dyk05}, \cite{Dyk051}
as the limit of slow modulation $\Omega << 1$ where "the driving
 frequency is small compared to the relaxation rate in the absence
 of fluctuations and the system remains in quasi-equilibrium" and by
the authors of the \cite{Leh00}, \cite{Leh000},
\cite{Leh03} theory as the one that "goes up to driving frequencies of
 the order of the inverse instanton time which is related to the
curvatures of the potential". In the
adiabatic regime the scaling behavior of the prefactor for the
 cubic potential is given by a
 simple analytical formula \cite{Dyk05}, \cite{Dyk051}.

However the existing literature leaves room for
 parallel activity for the following reasons. To attract attention
 of chemists and biochemists
to the modulated thermally activated escape theory within the Kramers'
approach
it is necessary to present its final results in as simple and
understandable form as is done, e.g., in the transition state theory.
 On the contrary the results of both \cite{Leh00},
 \cite{Leh000},
\cite{Leh03} and \cite{Sme99}, \cite{Sme99a}, \cite{Ryv04},
\cite{Dyk04}, \cite{Dyk05}, \cite{Dyk051}
theories are presented via involved
notions and values characterizing the optimal action
corresponding to the minimizing path. Correct making use of these results
requires profound comprehension of their physical content and
mastering in depth the methods involved in their deriving.
As a matter of fact people engaged in applications
(whom the author of the present manuscript belongs to) as a rule
 are concerned with much more
modest objective: how at a given combination from the parameter space
(noise intensity $D$, modulation amplitude $A$, modulation
frequency $\Omega$ and characteristics of the static potential
$U(x)$) to evaluate the escape rate enhancement in the presence of
driving at least for a simple analytically smooth (i.e., not
piecewise) metastable or bistable potential? That is why it is
desirable to have the period-average escape rate in
a closed analytical form
and explicitly expressed only via the parameters $D$, $A$, $\Omega$
and characteristics of the static potential $U(x)$
including no other physical values. In other words a theory convinient for
applications should restrict the physical
content of the resulting formula only by the notions used at
initial setting the problem.
Besides both \cite{Leh00}, \cite{Leh000}, \cite{Leh03} and
\cite{Sme99}, \cite{Sme99a}, \cite{Ryv04}, \cite{Dyk04},
\cite{Dyk05}, \cite{Dyk051} theories invoke to rather
sophisticated methods such as, e.g., path
integrals technique. The initial mathematical formulation of the
problem is a partial differential equation and it seems interesting
to see what results can be obtained among others by means of
 usual mathematics.
These reasons motivate the appearance of the present manuscript.
Our aim is to derive by means of elementary methods a closed
analytical form of the formula for the escape rate enhancement
in the low frequency limit for arbitrary values of
the modulation ampitude to noise intensity ratio $A/D$ except a
 narrow region near the bifurcation point.
Our approach is not based on a physical idea a priory inserted into the
theory but rather is a direct purely mathematical treatment of the problem.
We define the low frequency limit as the one
where there is no essential dependence on frequency so that the formal limit
$\Omega \rightarrow 0$ in the appropriate equations can be taken. At the
same time we observe the requirement
 $\Gamma_K << \Omega$ (where $\Gamma_K$ is the stationary
 Kramers' rate) that is necessary
for the
 efficient averaging over the period to be possible. The latter means
that all interesting phenomena related to the so called
stochastic resonance (taking place at $\Omega = \pi \Gamma_K$) are
 beyond the scope of the present theory. As the value
$\Gamma_K \propto exp \left [-\left(U_{max}-U_{min}\right)/D \right]$ is
usually vanishingly small there certainly should be a range
(perhaps $\Gamma_K << \Omega << D$)
 where the contradictory requirements  $\Gamma_K << \Omega$ and
$\Omega \rightarrow 0$ can be reconciled. As this range
turns out to be more restricted than that of
slow modulation
$\Omega << 1$ we use the term
low frequency limit instead of adiabatic regime to avoid confusion.

The manuscript is organized as follows. In Sec.2 the
problem is formulated and its solution is argued to be sought by
perturbation technique for the action in powers of the
modulation amplitude. Sec.3 and Sec.4 are devoted to the first and
second order contributions into action respectively.
 In Sec.5 the results are
combined in a closed
 form for the escape rate enhancement in the low frequency limit.
In Sec.6 the formula is used to obtain plots. In Sec. 7 a simple
analytical formula for the limiting
 case of moderately strong modulation
 $D << A << \sqrt D$ is obtained.
  In Sec.8 the results are discussed and the conclusions
are summarized. In the Appendix some technical
details are presented.\\

\section{Formulation of the problem}
\subsection {Setting the stage}
In this preliminary Sec. we pose the problem and remind some facts on the Kramers'
theory to introduce designations and notions used
further. In the Kramers' model a chemical reaction is considered as the
escape of a Brownian particle from the well of a
potential $U(x)$ along the reaction coordinate $x$ with $x_a$ being the point
of the bottom of the well and $x_b$ being the point of its top. The problem of interest
is to take into account the presence of periodic driving with modulation amplitude
$A$ and frequency $\Omega$.  For the driving we adopt without any
serious loss of generality the commonly used form
\begin{equation}
\label{eq1} f(t)=A sin(\Omega t)
\end{equation}
The results obtained for this simplest case can be directly
generalized to any arbitrary periodic driving because the latter
can be expanded into a Fourier series. In the overdamped limit
(strong friction case) the Fokker-Planck equation (FPE) for the
probability distribution function $P(x,t)$ is
\begin{equation}
\label{eq2} \dot P(x,t)=-F'(x)P(x,t)-\biggl[F(x)+
f(t)\biggr]P'(x,t)+DP^{\prime\prime}(x,t)
\end{equation}
where the dot denotes a derivative in time, the prime denotes a
derivative in coordinate, $F(x)=-U'(x)$ is the time independent
force field and $D$ is the so called noise intensity that is
actually the ratio of temperature in energetic units to the
barrier height. In the absence of periodic driving ($f(t)=0$) the
stationary limit of (\ref{eq2}) is
\begin{equation}
\label{eq3} 0=\dot P(x,t)=-\Bigl[F(x)P(x,t)-DP'(x,t)\Bigr]'
\end{equation}
which can be integrated to yield for the stationary (Kramers')
probability distribution function $P_K(x)$ the equation
\begin{equation}
\label{eq4} J=-DP'_K(x)+F(x)P_K(x)
\end{equation}
where $J$ is the stationary flux.
If we adopt the boundary condition as the absorbtion
 $P_K(x_c)=0$ at some point $x_c$ (with $x_c > x_b$ where $x_b$
 is the barrier top) then we have
\begin{equation}
\label{eq5} P_K(x)=\frac{J}{D}exp\Bigl(-U(x)/D\Bigr)
\int\limits_{x}^{x_c}dy\ exp\Bigl(U(y)/D\Bigr)
\end{equation}
For the Kramers' rate (taking into account that
 $N=\int\limits^{x_b}_{-\infty}dx\ P_K(x)\approx 1$) we have
\begin{equation}
\label{eq6} \Gamma_K=\frac {J}{N}\approx J
\approx \frac{\omega_a \omega_b}{2\pi}exp\Bigl[-\Bigl(U(x_b)-U(x_a)\Bigr)/D\Bigr]
\end{equation}
where $\omega_a=\sqrt{U^{\prime\prime}(x_a)}$ and
 $\omega_b=\sqrt{\vert U^{\prime\prime}(x_b)\vert}$.

In the presence of driving the position of the barrier top becomes
time dependent ($q_b(t)\approx x_b-f(t)$) and the population of the
well is
\begin{equation}
\label{eq7} N(t) =\int\limits_{-\infty}^{q_b(t)}dx \ P(x,t)
\end{equation}
A convenient operational definition of the reaction rate constant
(adopted, e.g., in  \cite{Leh00}, \cite{Leh000} and
 \cite{Leh03}) is
\begin{equation}
\label{eq8} \Gamma(t) =-\frac{\dot N(t)}{N(t)}
\end{equation}
It should be stressed that the
behavior of this value is sensitive to the
actual choice of the absorbtion boundary $x_c(t)$ where
 we set $P(x_c(t),t)=0$. The requirement we adopt further is that the
absorbtion point should be sufficiently far from the barrier top
\begin{equation}
\label{eq9} x_c(t)-x_b(t) >> A
\end{equation}
This requirement is in accordance with that to measure the current
well behind the boundary argued in \cite{Dyk051}.

Let us consider, e.g., the the simplest cubic (metastable) potential
 $U(x)=x^2/2-x^3/3$ (CP). In this case $x_a=0$, $x_b=1$,
 $\omega_b=1$, $\omega_a=1$ and
$q_b(t)\approx 1-A sin(\Omega t)$,
$q_a(t)\approx A sin(\Omega t)$. By analogy we adopt
for the general case
\begin{equation}
\label{eq10} q_b(t)\approx x_b-A sin(\Omega t);\ \ \ \
\dot q_b(t)=-\dot f(t);
\ \ \ \ q_a(t)\approx x_a+A sin(\Omega t)
\end{equation}
For the quartic (bistable) potential  $U(x)=-x^2/2+x^4/4$ (QP) we
 also have
$\omega_b=1$ while $x_a=-1$, $x_b=0$ and $\omega_a=\sqrt 2$.

Thus taking into account  that $N(t)\approx 1$ and (\ref{eq2}) we obtain
from (\ref{eq8})
\begin{equation}
\label{eq11} \Gamma(t)\approx - DP'(q_b(t),t)+
[F(q_b(t))+f(t)+\dot f(t)]P(q_b(t),t)
\end{equation}
The rate constant averaged over the period of oscillations
$T=2\pi/\Omega$ is
\begin{equation}
\label{eq12} \overline\Gamma=\frac{1}{T}
\int\limits_{t}^{t+T}ds \ \Gamma(s)
\end{equation}
Our final aim is to calculate the escape rate enhancement
\[
\Delta\equiv
\frac{\overline\Gamma}{\Gamma_K}\approx\frac{1}{T J}
\int\limits_{t}^{t+T}ds\ [F(q_b(s))+f(s)+\dot f(s)]P(q_b(s),s)-
\]
\begin{equation}
\label{eq13} \frac{D}{T J}
\int\limits_{t}^{t+T}ds\ P'(q_b(s),s)
\end{equation}
To attain this goal we will also need the probability distribution function near the bottom of
the well that is known to be \cite{Lud75}, \cite{Dyk051}
\begin{equation}
\label{eq14} P(x,t)\approx \frac{1}{\sqrt{2\pi D \sigma_a^2(t)}}
\exp \left\{-\frac{\left[x-q_a(t)\right]^2}{2D\sigma_a^2(t)}\right\}
\end{equation}
Here $\sigma_a(t)$ is the dispersion that can be identified with the
inverse frequency of the well $\sigma_a(t)=1/\omega_a(t)$. The latter can
be evaluated, e.g., for the CP $\omega_a(t)\approx 1-f(t)$.
By analogy we adopt for the general case
\begin{equation}
\label{eq15} \omega_a(t)\approx \omega_a-f(t)
\end{equation}
Substituting (\ref{eq11}) and (\ref{eq15}) into (\ref{eq14}) we obtain
\begin{equation}
\label{eq16} P(x_a,t)\approx \frac{\omega_a-A sin(\Omega t)}
{\sqrt{2\pi D }}
\exp \left\{-\frac{\left[A sin(\Omega t)
\left(\omega_a-A sin(\Omega t)\right)\right]^2}{2D}\right\}
\end{equation}

\subsection {Form of the action}
We seek the solution of (\ref{eq2}) in the form
\begin{equation}
\label{eq17} P(x,t)=P_K(x)u(x,t)
\end{equation}
The form (\ref{eq17}) means that
\begin{equation}
\label{eq18} P(x_c,t)=0
\end{equation}
 because
$P_K(x_c)=0$, i.e., adopting this form we have to neglect the
possible dependence of the absorbtion point $x_c(t)$ on time in
the presence of driving and assume $x_c(t)=x_c$ where $x_c$ is
that in the absence of driving. The latter may be justified by the
requirement (\ref{eq9}). In the present approach we adopt
this approximation without further discussing its validity.
Substitution of (\ref{eq17}) into (\ref{eq13}) with taking into
account (\ref{eq5}) yields for the value of interest
\[
\Delta=\frac{1}{T}
\int\limits_{t}^{t+T}ds\ u(q_b(s),s)+\frac{1}{T D}
\int\limits_{t}^{t+T}ds\ exp\Bigl(-U(q_b(s))/D\Bigr)\times
\]
\begin{equation}
\label{eq19}
\int\limits_{q_b(s)}^{x_c}dy\ exp\Bigl(U(y)/D\Bigr)
\Bigl[[f(s)+\dot f(s)]u(q_b(s),s)-Du'(q_b(s),s)\Bigr]
\end{equation}
We denote
\begin{equation}
\label{eq20} \Psi (x)=\frac{P'_K(x)}{P_K(x)}
\end{equation}
For the function $u(x,t)$ we obtain the equation
\[
 \dot u(x,t)=Du^{\prime\prime}(x,t)-
\]
\begin{equation}
\label{eq21}
\biggl[F(x)+f(t)-2D\Psi (x)\biggr]u'(x,t)-f(t)\Psi (x) u(x,t)
\end{equation}
At $A=0$ we must have $u(x,t)\equiv 1$ for asymptotically large
time that suggests to seek the solution of (\ref{eq21}) in the form
\begin{equation}
\label{eq22}  u(x,t)= exp \bigl[ A \alpha (x,t)\bigr]
\end{equation}
This form can not be exact because in the
limit $A\rightarrow
0$ we obtain $P(x,t)=P_K(x)$ that can be valid only asymptotically at
$t\rightarrow\infty$.
We denote
\begin{equation}
\label{eq23} \Phi(x)=2D\Psi(x)-F(x)
\end{equation}
For the function $\alpha (x,t)$ we obtain the equation
\[
 \dot \alpha(x,t)=D\alpha^{\prime\prime}(x,t)+
DA\bigl[\alpha'(x,t)\bigr]^2+
\]
\begin{equation}
\label{eq24}
\biggl[\Phi(x)-A sin(\Omega t) \biggr]\alpha'(x,t)-
\Psi (x) sin(\Omega t)
\end{equation}

In the present manuscript we argue the point of view that in
the low frequency limit the function $\alpha (x,t)$ can be sought
as a perturbation series in the modulation amplitude $A$
\begin{equation}
\label{eq25} \alpha(x,t)=\varphi(x,t)+
A\chi(x,t)+A^2\mu(x,t)+O(A^3)
\end{equation}
The function $\alpha (x,t)$ by its final result
in (\ref{eq13}) plays
the same role as the action from the theories \cite{Sme99},
\cite{Ryv04}, \cite{Dyk05}, \cite{Dyk051} and \cite{Leh00},
 \cite{Leh000},
\cite{Leh03}. That is why we will also use this name.

The results of the present manuscript testify that at least in the low
frequency limit $\alpha' \propto 1/D$ so that
 $u'(x,t)\sim A/D\ u(x,t)$. Thus taking
into account that $f(t) \propto A$ and
 $\int\limits_{q_b(s)}^{x_c}dy \ exp\Bigl(U(y)/D\Bigr)
\sim \sqrt{D}exp\Bigl(-U(x_b)/D\Bigr)$
(see below) we generally have for the escape rate enhancement
\begin{equation}
\label{eq26} \Delta\approx\frac{1}{T}
\int\limits_{t}^{t+T}ds\ u(q_b(s),s)
\Biggl[1+O\Biggl(\frac{A}{\sqrt{D}}\Biggr)\Biggr]
\end{equation}
Though at large ratios $A/D$ the term
$O\bigl(\frac{A}{\sqrt{D}}\bigr) \sim 1$ we will not take it into account
in the present paper despite of the fact that the method of calculation
developed below enables us to treat it. This term seems to give minor
correction and its role will be considered elsewhere.

Substituting (\ref{eq25}) in (\ref{eq22}), (\ref{eq21}) and collecting
 the terms at powers of $A$ we obtain the
 following system of equations
\begin{equation}
\label{eq27} \dot \varphi(x,t)=D\varphi^{\prime\prime}(x,t)+
\Phi(x)\varphi'(x,t)-\Psi(x)\ sin(\Omega t)
\end{equation}
\begin{equation}
\label{eq28} \dot \chi(x,t)=D\chi^{\prime\prime}(x,t)+
 \Phi(x)\chi'(x,t)+
D\Bigl[\varphi'(x,t)\Bigr]^2-
sin(\Omega t)\ \varphi'(x,t)
\end{equation}
\begin{equation}
\label{eq29} \dot \mu(x,t)=D\mu^{\prime\prime}(x,t)+
\Phi(x)\mu'(x,t)+\Bigl[2D\varphi'(x,t)-
sin(\Omega t)\Bigr]\chi'(x,t)
\end{equation}
etc.
From (\ref{eq2}), (\ref{eq18}) and (\ref{eq21}) we obtain the boundary
conditions
\begin{equation}
\label{eq30} \varphi'(x_c,t)=\frac{1}{2D}sin(\Omega t);\ \ \ \ \ \ \
\chi'(x_c,t)=0;\ \ \ \ \ ...
\end{equation}
From (\ref{eq16}) we obtain
\begin{equation}
\label{eq31} \varphi(x_a,t)=-\frac{1}{\omega_a}sin(\Omega t);\ \ \ \ \ \ \
\chi(x_a,t)=-\frac{\omega_a^2}{2D} sin^2(\Omega t);\ \ \ \ \ ...
\end{equation}

\section{First order contribution into action}
\subsection {Equations for the first order contribution into action}
We start from (\ref{eq27}). Its solution for asymptotically large time
can be sought in the form
\begin{equation}
\label{eq32} \varphi(x,t)=g(x)sin(\Omega t)+h(x)cos(\Omega t)
\end{equation}
From (\ref{eq30}) and (\ref{eq31}) we obtain
the boundary conditions
\begin{equation}
\label{eq33} g'(x_c)=\frac{1}{2D};\ \ \ \ \ \ h'(x_c)=0
\end{equation}
\begin{equation}
\label{eq34}  g(x_a)=-\frac{1}{\omega_a};\ \ \ \ \ \ h(x_a)=0
\end{equation}
Substitution of (\ref{eq32}) into (\ref{eq27}) yields a system of
coupled equations
\begin{equation}
\label{eq35} Dg^{\prime\prime}(x)+\Phi g'(x)+\Omega h(x)=\Psi(x)
\end{equation}
\begin{equation}
\label{eq36} Dh^{\prime\prime}(x)+\Phi h'(x)-\Omega g(x)=0
\end{equation}
This system is rather difficult for analytical treatment.
However in the
present manuscript we restrict ourselves by the low frequency limit
$\Omega \rightarrow 0$.
 In this case
 the equations (\ref{eq35}), (\ref{eq36})) are decoupled
\begin{equation}
\label{eq37} g^{\prime\prime}(x)+\frac{\Phi}{D} g'(x)=
\frac{\Psi(x)}{D}
\end{equation}
\begin{equation}
\label{eq38} h^{\prime\prime}(x)+\frac{\Phi}{D} h'(x)=0
\end{equation}

Taking into account that
\begin{equation}
\label{eq39} exp
\Biggl[ \frac{1}{D}\int\limits_{z}^{s}dr\ \Phi(r)\Biggr]=
\frac{P^2_K(s)}{P^2_K(z)}exp\Biggl[ \frac{U(s)-U(z)}{D}\Biggr]
\end{equation}
we obtain the solution of (\ref{eq37}), (\ref{eq38}) satisfying
(\ref{eq33}), (\ref{eq34}) as
\begin{equation}
\label{eq40}g(x)=-\frac{1}{\omega_a}-\frac{1}{D}
\int\limits_{x_a}^{x}dy\ \frac{exp\Bigl(-\frac{U(y)}{D}\Bigr)}
{P^2_K(y)}
\int\limits_{y}^{x_c}dz\ P^2_K(z)exp\Biggl(\frac{U(y)}{D}\Biggr)
\Psi(z)
\end{equation}
\begin{equation}
\label{eq41}\ \ \ \ \ \ \ \ \ \ \ \ \ h(x)\equiv 0;
\ \ \ \ \ \ \ \ \ \ \ \ \ \ \ \ \ \ \
h'(x)\equiv 0
\end{equation}

We substitute $P_K(x)$ from (\ref{eq5}) into (\ref{eq40}) and
notice that
\begin{equation}
\label{eq42} dy \frac{exp\left(\frac{U(y)}{D}\right)}
{\left[ \int\limits_{y}^{x_c}ds\ exp\left(\frac{U(s)}{D}\right)\right]^2}=
d \frac{1}{\int\limits_{y}^{x_c}ds\ exp\left(\frac{U(s)}{D}\right)}
\end{equation}
Making use of integration by part and denoting
\begin{equation}
\label{eq43} \Gamma_0(x)=\int\limits_{x}^{x_c}ds\ exp\left(\frac{U(s)}
{D}\right)
\end{equation}
\begin{equation}
\label{eq44} E_1(x)=\int\limits_{x}^{x_c}dz\ \Psi(z)exp\left(-\frac{U(z)}
{D}\right)\left[\int\limits_{z}^{x_c}ds\ exp\left(\frac{U(s)}
{D}\right) \right]^2
\end{equation}
\begin{equation}
\label{eq45} E_2(x)=\int\limits_{x_a}^{x}dz\ \Psi(z)exp\left(-\frac{U(z)}
{D}\right)\int\limits_{z}^{x_c}ds\ exp\left(\frac{U(s)}
{D}\right)
\end{equation}
we obtain
\begin{equation}
\label{eq46} g(x)=-\frac{1}{\omega_a}-\frac{1}{D}\Biggl\lbrace
\frac{1}{\Gamma_0(x)}E_1(x)-
\frac{1}{\Gamma_0(x_a)}E_1(x_a)+
E_2(x)
 \Biggr\rbrace
\end{equation}

Making use of (\ref{eq20}) and (\ref{eq5}) we obtain
\begin{equation}
\label{eq47} E_1(x)=-exp\left(-\frac{U(x)}
{D}\right)\Gamma_0^2(x)+\int\limits_{x}^{x_c}ds\ exp\left(\frac{U(s)}
{D}\right)(s-x)
\end{equation}
\begin{equation}
\label{eq48} E_2(x)=exp\left(-\frac{U(x)}
{D}\right)\Gamma_0(x)-exp\left(-\frac{U(x_a)}
{D}\right)\Gamma_0(x_a)
\end{equation}
We recall (\ref{eq11}) and denote
\[
S(q_b(t))=\frac{1}{\Gamma_0(q_b(t))}
\int\limits_{q_b(t)}^{x_c}ds\ exp\left(\frac{U(s)}
{D}\right)(s-q_b(t))-
\]
\begin{equation}
\label{eq49} \frac{1}{\Gamma_0(x_a)}
\int\limits_{x_a}^{x_c}ds\ exp\left(\frac{U(s)}
{D}\right)(s-x_a)
\end{equation}
\begin{equation}
\label{eq50} \lambda=\frac{\omega_b^2}{2D}
\end{equation}
\begin{equation}
\label{eq51} Q_n(x)=
\int\limits_{x}^{x_c}ds\ exp\left[-\lambda (s-x_b)^2\right](s-x)^n
\end{equation}
Making use of N2.3.15.1 from \cite{Pr81} we obtain
\[
 Q_n(q_b(t))\approx
n!\left(2\lambda\right)^{-\frac{n+1}{2}}\times
\]
\begin{equation}
\label{eq52}
\exp\left[-\frac{\lambda A^2 sin^2 (\Omega t)}{2} \right]
D_{-(n+1)}\left[-\sqrt{2\lambda}\ A sin (\Omega t) \right]
\end{equation}
where $D_n(x)$ is a parabolic cylinder function. Making use of its
known properties we have
\begin{equation}
\label{eq53}  Q_0(q_b(t))= \frac{\sqrt \pi}{2\sqrt \lambda}
 erfc\left[-\sqrt{\lambda}\ A sin (\Omega t) \right]
\end{equation}
\begin{equation}
\label{eq54} \Gamma_0\left(q_b(t)\right)\approx
\frac{\sqrt{\pi D}}{\sqrt{2}\omega_b}exp\left(\frac{U(x_b)}{D}\right)
erfc\left(-\frac{\omega_b A sin(\Omega t)}{\sqrt{2D}} \right)
\end{equation}
\[
Q_1(q_b(t))= \frac{\sqrt \pi}{2\sqrt \lambda}
 A sin (\Omega t)
 erfc\left[-\sqrt{\lambda}\ A sin (\Omega t) \right]+
\]
\begin{equation}
\label{eq55}
\frac{1}{2\lambda}exp\left(-\lambda A^2
sin^2\left(\Omega t\right)\right)
\end{equation}
where $erfc(x)$ is an additional error function.
 Making use of (\ref{eq53}) and (\ref{eq55}) we obtain
(see Appendix for details)
\begin{equation}
\label{eq56} S(q_b(t))\approx -(q_b(t)-x_a)+ O(\sqrt D)
\end{equation}
and
\begin{equation}
\label{eq57}
g(q_b(t))\approx\frac{q_b(t)-x_a}{D}-\frac{1}{\omega_a}
\end{equation}
The value $\frac{1}{\omega_a}$ is negligibly small compared with
$\frac{x_b-x_a}{D}$ and can be omitted.
Thus finally we obtain the first order contribution into action as
\begin{equation}
\label{eq58} \varphi(q_b(t),t) \approx
\Biggl[\frac{x_b-A sin(\Omega t)-x_a}{D}\Biggr] sin (\Omega t)
\end{equation}

\section{Second order contribution into action}
\subsection{Equations for the second order contribution into action}
From (\ref{eq28}) and the results for the first order contribution into
action we obtain
\begin{equation}
\label{eq59} \dot \chi(x,t)=D\chi^{\prime\prime}(x,t)+
 \Phi(x)\chi'(x,t)+L(x)\bigl[1-cos(2\Omega t)\bigr]
\end{equation}
where we denote
\begin{equation}
\label{eq60} L(x)=\frac{g'(x)\bigl[Dg'(x)-1\bigr]}{2}
\end{equation}
We seek the solution of (\ref{eq59}) in the form
\begin{equation}
\label{eq61} \chi(x,t)=v(x)+\phi(x,t)
\end{equation}
where the new functions obey the equations
\begin{equation}
\label{eq62} Dv^{\prime\prime}(x)+\Phi(x)v'(x)=-L(x)
\end{equation}
\begin{equation}
\label{eq63} \dot \phi(x,t)=D\phi^{\prime\prime}(x,t)+
 \Phi(x)\phi'(x,t)-L(x)cos(2\Omega t)
\end{equation}
From (\ref{eq30}) and (\ref{eq31}) we obtain the boundary conditions
\begin{equation}
\label{eq64} v(x_a)=-\frac{\omega_a^2}{4D}; \ \ \ \ \ \ \
\ \ \ \ \ \ \ \ v'(x_c)=0
\end{equation}
\begin{equation}
\label{eq65} \phi(x_a,t)=\frac{\omega_a^2}{4D}cos(2\Omega t); \ \ \
\ \ \ \ \ \ \ \ \ \ \phi'(x_c,t)=0
\end{equation}

\subsection{$v(x)$ function}
The solution for the function $v(x)$ is
\begin{equation}
\label{eq66} v(x)=-\frac{\omega_a^2}{4D}+\frac{1}{D}
\int\limits_{x_a}^{x}dy\ \frac{exp\Bigl(-U(y)/D\Bigr)}{P_K^2(y)}
\int\limits_{y}^{x_c}dz\ L(z)P_K^2(z)exp\Bigl(U(z)/D\Bigr)
\end{equation}
We denote
\begin{equation}
\label{eq67} I(x)=-\int\limits_{x}^{x_c}dz\ L(z)
exp\left(-\frac{U(z)}
{D}\right)\int\limits_{z}^{x_c}ds\ exp\left(\frac{U(s)}
{D}\right)\int\limits_{x}^{z}dy\ exp\left(\frac{U(y)}
{D}\right)
\end{equation}
After straightforward calculations we obtain
\begin{equation}
\label{eq68} v(q_b(t))=-\frac{\omega_a^2}{4D}+
\frac{1}{D}\left\{\frac{I\left(q_b(t)\right)}
{\Gamma_0\left(q_b(t)\right)}-
\frac{I\left(x_a\right)}{\Gamma_0\left(x_a\right)}\right\}
\end{equation}
The inner integrals for both $I\left(q_b(t)\right)$ and
 $I\left(x_a\right)$ mave maximum at $z\approx x_b$. That is why
we adopt the following approximations
\begin{equation}
\label{eq69} I\left(q_b(t)\right)\approx -L\left(q_b(t)\right)\Theta
\end{equation}
\begin{equation}
\label{eq70} I(x_a)\approx -L(x_b)\Lambda
\end{equation}
where
\begin{equation}
\label{eq71} \Theta=\int\limits_{q_b(t)}^{x_c}dz\
exp\left(-\frac{U(z)}
{D}\right)\int\limits_{z}^{x_c}ds\ exp\left(\frac{U(s)}
{D}\right)\int\limits_{q_b(t)}^{z}dy\ exp\left(\frac{U(y)}
{D}\right)
\end{equation}
\begin{equation}
\label{eq72} \Lambda=\int\limits_{x_a}^{x_c}dz\
exp\left(-\frac{U(z)}
{D}\right)\int\limits_{z}^{x_c}ds\ exp\left(\frac{U(s)}
{D}\right)\int\limits_{x_a}^{z}dy\ exp\left(\frac{U(y)}
{D}\right)
\end{equation}
Denoting
\begin{equation}
\label{eq73} H=\frac{\omega_a^2}{4}-\frac{\Lambda}
{\Gamma_0(x_a)}L(x_b)+
\frac{\Theta}{\Gamma_0\left(q_b(t)\right)}L\left(q_b(t)\right)
\end{equation}
 we obtain
\begin{equation}
\label{eq74} v(q_b(t))=-\frac{H}{D}
\end{equation}

\subsection{$\phi(x,t)$ function}
We seek the solution of (\ref{eq63}) in the form
\begin{equation}
\label{eq75} \phi(x,t)=r(x) sin(2\Omega t)+ s(x) cos(2\Omega t)
\end{equation}
For the new functions we obtain the equations
\begin{equation}
\label{eq76} Dr^{\prime\prime}(x)+\Phi r'(x)+2\Omega s(x)=0
\end{equation}
\begin{equation}
\label{eq77} Ds^{\prime\prime}(x)+\Phi s'(x)-2\Omega r(x)=L(x)
\end{equation}
Recalling that we are in the low frequency limit
 we decouple the equations (the limit
$\Omega \rightarrow 0$ in (\ref{eq76}), (\ref{eq77}))
\begin{equation}
\label{eq78} r^{\prime\prime}(x)+\frac{\Phi}{D} r'(x)=0
\end{equation}
\begin{equation}
\label{eq79} s^{\prime\prime}(x)+\frac{\Phi}{D} s'(x)=\frac{L(x)}{D}
\end{equation}
From (\ref{eq64}), (\ref{eq65}) we obtain the boundary conditions for these
equations
\begin{equation}
\label{eq80} r(x_a)=0; \ \ \ \ \ \ \
\ \ \ \ \ \ \ \ r'(x_c)=0
\end{equation}
\begin{equation}
\label{eq81} s(x_a)=\frac{\omega_a^2}{4D}; \ \ \
\ \ \ \ \ \ \ \ \ \ s'(x_c)=0
\end{equation}
The solutions satisfying these boundary conditions are
\begin{equation}
\label{eq82}\ \ \ \ \ \ \ \ \ \ \ \ \ s(x)\equiv -v(x)
\end{equation}
\begin{equation}
\label{eq83}\ \ \ \ \ \ \ \ \ \ \ \ \ r(x)\equiv 0;
\ \ \ \ \ \ \ \ \ \ \ \ \ \ \ \ \ \ \
r'(x)\equiv 0
\end{equation}
Thus we obtain
\begin{equation}
\label{eq84} s(q_b(t))=\frac{H}{D}
\end{equation}

\subsection{$\chi(x,t)$ function}
Combining the results we obtain the second order contribution into action
\begin{equation}
\label{eq85} \chi\left(q_b(t),t\right)\approx -\frac{H}{D}
\left(1-cos(2\Omega t)\right)
\end{equation}
To calculate $H$ from (\ref{eq73}) we need $L(x_b)$ and
 $L\left(q_b(t)\right)$. Taking into account
(\ref{eq60}) the latter means that we need $g'(x_b)$ and
 $g'\left(q_b(t)\right)$. From (\ref{eq40})
we have
\begin{equation}
\label{eq86} g'(x)=-\frac{E_1(x)}
{D \Gamma_0^2(x)}exp\left(\frac{U(x)}
{D}\right)
\end{equation}
From (\ref{eq47}) and (\ref{eq55}) we obtain
\begin{equation}
\label{eq87} E_1(x_b)=-\Gamma_0^2(x_b) exp\left(-\frac{U(x_b)}
{D}\right)+\frac{1}{2\lambda}exp\left(\frac{U(x_b)}
{D}\right)
\end{equation}
where
\begin{equation}
\label{eq88} \Gamma_0(x_b)\approx
 \frac{\sqrt{2\pi D}}{2\omega_b}exp\left(\frac{U(x_b)}
{D}\right)
\end{equation}
As a result we have
\begin{equation}
\label{eq89} g'(x_b)=\frac{\pi-2}{\pi D}
\end{equation}
Thus
\begin{equation}
\label{eq90} L(x_b)=-\frac{\pi-2}{\pi^2 D}
\end{equation}
From (\ref{eq54}) and (\ref{eq55}) we obtain
\[
g'\left(q_b(t)\right)=\frac{1}{D}\Biggl\lbrace 1-
\frac{2\omega_b^2}{\pi}
exp\left(\frac{U\left(q_b(t)\right)-U(x_b)}{D}\right)
/erfc^2\left(-\frac{\omega_b A sin(\Omega t)}{\sqrt{2D}}\right)\times
\]
\begin{equation}
\label{eq91} \left[\frac{\sqrt{2\pi}A sin(\Omega t)}
{2\omega_b \sqrt D}erfc\left(-\frac{\omega_b A sin(\Omega t)}
{\sqrt{2D}}\right)+\frac{1}{\omega_b^2}
exp\left(-\frac{\omega_b^2 A^2 sin^2(\Omega t)}
{2D}\right)
\right]\Biggr \rbrace
\end{equation}
Substitution of (\ref{eq91}) into (\ref{eq60}) yields
$L\left(q_b(t)\right)$ that is not written out explicitly to save room.

\subsection{$\Lambda$ value}
Let us evaluate the integral $\Lambda$ given by (\ref{eq72}). The maximum of
the inner integrals takes place at $z\approx x_b$. In this case both inner
integrals $\propto exp\left(\frac{U(x_b)}{D}\right)$, otherwise only one of them is such and the whole integral
should be smaller. Thus we approximate at $z\approx x_b$
\begin{equation}
\label{eq92}\int\limits_{z}^{x_c}ds\ exp\left(\frac{U(s)}
{D}\right)\approx \frac{\sqrt{2\pi D}}{2\omega_b}exp\left(\frac{U(z)}
{D}\right)
\end{equation}
\begin{equation}
\label{eq93}\int\limits_{x_a}^{z}dy\ exp\left(\frac{U(y)}
{D}\right)\approx \frac{\sqrt{2\pi D}}{2\omega_b}exp\left(\frac{U(z)}
{D}\right)
\end{equation}
As a result we obtain
\begin{equation}
\label{eq94}\Lambda\approx
\frac{\left(2\pi D\right)^{3/2}}{4\omega_b^3}exp\left(\frac{U(x_b)}
{D}\right)
\end{equation}
Taking into account that
\begin{equation}
\label{eq95} \Gamma_0(x_a)\approx
\frac{\sqrt{2\pi D}}{\omega_b}exp\left(\frac{U(x_b)}
{D}\right)
\end{equation}
we obtain
\begin{equation}
\label{eq96} \frac{\Lambda}{\Gamma_0(x_a)}\approx \frac{\pi D}{2\omega_b^2}
\end{equation}

\subsection{$\Theta$ value}
Now we start rather tedious evaluation of the function $\Theta$ given
 by (\ref{eq71}). Let us first calculate the auxiliary value $\Theta_0$
\begin{equation}
\label{eq97} \Theta_0=\int\limits_{x_b}^{x_c}dz\
exp\left(-\frac{U(z)}
{D}\right)\int\limits_{z}^{x_c}ds\ exp\left(\frac{U(s)}
{D}\right)\int\limits_{x_b}^{z}dy\ exp\left(\frac{U(y)}
{D}\right)
\end{equation}
The inner integrals have maximum at $z\approx x_b$. In this case we expand
the last integral into a Taylor series
\begin{equation}
\label{eq98} \int\limits_{x_b}^{z}dy\ exp\left(\frac{U(y)}
{D}\right)\approx (z-x_b)exp\left(\frac{U(x_b)}
{D}\right)\approx (z-x_b)exp\left(\frac{U(z)}
{D}\right)
\end{equation}
Then
\begin{equation}
\label{eq99} \Theta_0\approx \frac{\sqrt{2\pi D}}{2\omega_b}Q_1(x_b)
\end{equation}
Making use of (\ref{eq55}) we obtain
\begin{equation}
\label{eq100} \Theta_0\approx
\frac{\left(2\pi D\right)^{3/2}}{4\pi\omega_b^3}exp\left(\frac{U(x_b)}
{D}\right)
\end{equation}
To calculate $\Theta$ we decompose it as follows
\begin{equation}
\label{eq101} \Theta=\Theta_0+W_1+W_2+W_3
\end{equation}
where we denote
\begin{equation}
\label{eq102} W_1=\left[\int\limits_{q_b(t)}^{x_b}dy\ exp\left(\frac{U(y)}
{D}\right)\right]\int\limits_{x_b}^{x_c}dz\
exp\left(-\frac{U(z)}
{D}\right)\int\limits_{z}^{x_c}ds\ exp\left(\frac{U(s)}
{D}\right)
\end{equation}
\begin{equation}
\label{eq103} W_2=\int\limits_{q_b(t)}^{x_b}dz\
exp\left(-\frac{U(z)}
{D}\right)\int\limits_{z}^{x_c}ds\ exp\left(\frac{U(s)}
{D}\right)\int\limits_{x_b}^{z}dy\ exp\left(\frac{U(y)}
{D}\right)
\end{equation}
\begin{equation}
\label{eq104} W_3=\left[\int\limits_{q_b(t)}^{x_b}dy\ exp\left(\frac{U(y)}
{D}\right)\right]\int\limits_{q_b(t)}^{x_b}dz\
exp\left(-\frac{U(z)}
{D}\right)\int\limits_{z}^{x_c}ds\ exp\left(\frac{U(s)}
{D}\right)
\end{equation}
We begin with the itegral in the square brackets from $W_1$ and $W_3$.
It can be evaluated as follows
\[
exp\left(-\frac{U(x_b)}{D}\right)\int\limits_{q_b(t)}^{x_b}dy\ exp\left(\frac{U(y)}
{D}\right)\approx
\int\limits_{q_b(t)}^{x_b}dy\ exp\left(-\lambda (y-x_b)^2\right)=
\]
\begin{equation}
\label{eq105}
\int\limits_{-A sin(\Omega t)}^{0}ds\ exp\left(-\lambda s^2\right)=
\frac{\sqrt{\pi}}{2\sqrt{\lambda}}
erf\left(\sqrt{\lambda}A sin(\Omega t) \right)
\end{equation}
Then $W_1$ can be approximated as
\begin{equation}
\label{eq106} W_1\approx \frac{\sqrt{\pi}}{2\sqrt{\lambda}}
erf\left(\sqrt{\lambda}A sin(\Omega t) \right)
\int\limits_{x_b}^{x_c}dz\
\int\limits_{z}^{x_c}ds\ exp\left(\frac{U(s)}
{D}\right)
\end{equation}
Altering the order of integration in the double integral we obtain
\[
\int\limits_{x_b}^{x_c}dz\
\int\limits_{z}^{x_c}ds\ exp\left(\frac{U(s)}
{D}\right)=
\]
\begin{equation}
\label{eq107} \int\limits_{x_b}^{x_c}ds\ exp\left(\frac{U(s)}
{D}\right)(s-x_b)\approx exp\left(\frac{U(x_b)}{D}\right)Q_1(x_b)
\end{equation}
Making use of (\ref{eq55}) we obtain
\begin{equation}
\label{eq108} W_1\approx \Theta_0\
erf\left(\frac{\omega_b A sin(\Omega t)}{\sqrt{2D}} \right)
\end{equation}
At evaluation of the inner integrals in $W_2$ we apply the same arguments as
those used for deriving (\ref{eq98}). Then we have
\begin{equation}
\label{eq109} W_2\approx \frac{\sqrt{2\pi D}}{2\omega_b}
\int\limits_{q_b(t)}^{x_b}dz\ exp\left(\frac{U(z)}
{D}\right)(z-x_b)
\end{equation}
After evaluation of the integral we obtain
\begin{equation}
\label{eq110} W_2 \approx \Theta_0
\left[exp\left(-\frac{\omega_b^2 A^2 sin^2(\Omega t)}
{2D}\right)-1\right]
\end{equation}
For $W_3$ after taking into account (\ref{eq106}) we have the approximation
\begin{equation}
\label{eq111} W_3 \approx \frac{\sqrt{\pi}}{2\sqrt{\lambda}}
erf\left(\sqrt{\lambda}A sin(\Omega t) \right)
\int\limits_{q_b(t)}^{x_b}dz\
\int\limits_{z}^{x_c}ds\ exp\left(\frac{U(s)}
{D}\right)
\end{equation}
For the double integral we have
\begin{equation}
\label{eq112}
\int\limits_{q_b(t)}^{x_b}dz\
\int\limits_{z}^{x_c}ds\ exp\left(\frac{U(s)}
{D}\right) \approx
 \frac{\sqrt{\pi}exp\left(\frac{U(x_b)}{D}\right)}{2\sqrt{\lambda}}
\int\limits_{-A sin(\Omega t)}^{0}dv\ erfc(\sqrt{\lambda}v)
\end{equation}
Making use of N1.5.1.9 from \cite{Pr83} we obtain
\[
 W_3\approx \frac{\pi D}{2\omega_b^2}exp\left(\frac{U(x_b)}{D}\right)
erf\left(\frac{\omega_b A sin(\Omega t)}{\sqrt{2D}} \right)
\Biggl\lbrace A sin(\Omega t)\times
\]
\begin{equation}
\label{eq113}erfc\left(-\frac{\omega_b A sin(\Omega t)}
{\sqrt{2D}} \right)+
\frac{\sqrt{2\pi D}}{\pi \omega_b}
\left[exp\left(-\frac{\omega_b^2 A^2 sin^2(\Omega t)}
{2D}\right)-1\right]
\Biggr\rbrace
\end{equation}
Combining the results we finally obtain
\[
\frac{\Theta}{\Gamma_0\left(q_b(t)\right)} \approx
\frac{D}{\omega_b^2}
\Biggl\lbrace
exp\left(-\frac{\omega_b^2 A^2 sin^2(\Omega t)}
{2D}\right)+
\]
\begin{equation}
\label{eq114}
\frac{\pi \omega_b A sin(\Omega t)}{\sqrt{2\pi D}}
erf\left(\frac{\omega_b A sin(\Omega t)}
{\sqrt{2D}} \right)\Biggr\rbrace
\end{equation}

\section{Result for the escape rate enhancement}
Combining the results and taking into account that
$T=\frac{2\pi}{\Omega}$ we finally obtain the escape rate
enhancement in the low frequency limit
\[
\Delta\approx\frac{1}{2\pi}
\int\limits_{0}^{2\pi}d\phi\ exp\Biggl\lbrace \frac{A}{D}
\Bigl[x_b-A sin(\phi)-x_a \Bigr] sin (\phi)-
\]
\[
\frac{2 A^2 sin^2(\phi)}{D}\Biggl\lbrace
\frac{\omega_a^2}{4}+\frac{\pi-2}{2\pi \omega_b^2}-
\frac{1}{\pi}\Biggl[
exp\left(-\frac{\omega_b^2 A^2 sin^2(\phi)}
{2D}\right)+
\]
\[
\frac{\pi \omega_b A sin(\phi)}{\sqrt{2\pi D}}
erf\left(\frac{\omega_b A sin(\phi)}
{\sqrt{2D}} \right)\Biggr]
exp\left(\frac{U\left(x_b-A sin(\phi)\right)-U(x_b)}{D}\right)\times
\]
\[
\Biggl\lbrace 1-
\frac{2\omega_b^2}{\pi}
exp\left(\frac{U\left(x_b-A sin(\phi)\right)-U(x_b)}{D}\right)
/erfc^2\left(-\frac{\omega_b A sin(\phi)}{\sqrt{2D}}\right)\times
\]
\[
 \left[\frac{\sqrt{2\pi}A sin(\phi)}
{2\omega_b \sqrt D}erfc\left(-\frac{\omega_b A sin(\phi)}
{\sqrt{2D}}\right)+\frac{1}{\omega_b^2}
exp\left(-\frac{\omega_b^2 A^2 sin^2(\phi)}
{2D}\right)
\right]\Biggr \rbrace\times
\]
\[
\Biggl[\frac{\sqrt{2\pi}A sin(\phi)}
{2\omega_b\sqrt D}erfc\left(-\frac{\omega_b A sin(\phi)}
{\sqrt{2D}}\right)+
\]
\begin{equation}
\label{eq115}\frac{1}{\omega_b^2}
exp\left(-\frac{\omega_b^2 A^2 sin^2(\phi)}
{2D}\right)
\Biggr]/erfc^2\left(-\frac{\omega_b A sin(\phi)}
{\sqrt{2D}}\right)\Biggr\rbrace\Biggr\rbrace
\end{equation}
This formula is the first result of the present manuscript.

\subsection{Plots}
The formula (\ref{eq115}) notwithstanding it looks very cumbersom can be
easily treated by, e.g., {\it Mathematica}. Prior doing it we recall that
we want to stay within the so called subthreshold driving regime.
The latter provides that the potential surface
always has a minimum and a maximum, i.e., the oscillating field is small
enough not to distort the physical picture of the chemical
reaction as the Brownian particle escape from the metastable
state. This regime is defined by the requirement $A \le A_c$ where
$A_c =0.25$ for
the case of the cubic (metastable) potential $U(x)=x^2/2-x^3/3$\ (CP)
and
$A_c = 2/(3\sqrt 3)\approx 0.4$ (see, e.g., \cite{Tal99} ) for
the case of the quartic (bistable) potential $U(x)=-x^2/2+x^4/4$\ (QP).
 In Fig.1 and Fig.2 the results
for the CP and QP respectively at relatively large values of the noise intensity
 are depicted. At the value $D = 5 \cdot 10^{-2}$ we have
$A_c/D \approx 5$ and $A_c/D \approx 8$ for the case of the CP and QP
respectively.  In Fig.3 and Fig.4 the results
for the CP and QP respectively at relatively small values of the noise intensity
 aredepicted. At the value $D = 3\cdot 10^{-3}$ we have
$A_c/D \approx 83$ and $A_c/D \approx 130$ for the cases of the CP and QP
respectively. These values limit the horizontal coordinate in the plots.

\subsection{Limit $1 << \frac{A}{D} << \frac{1}{\sqrt {D}}$}
Let us consider the limiting case of  moderately
strong modulation $D << A << \sqrt {D}$. We denote
\begin{equation}
\label{eq116} p =
1+\frac{2}{\omega_b^2}\Biggl[
\frac{\omega_a^2}{4}+\frac{\pi-2}{\pi}
\Biggl( \frac{1}{2\omega_b}-\frac{1}{\pi}\Biggr)
\Biggr ]
\end{equation}
Taking into account the requirement $A/\sqrt D << 1$ and
 discarding in (\ref{eq115}) the terms $O(A/\sqrt D)$ we obtain
\begin{equation}
\label{eq117} \Delta \approx \frac{1}{2\pi}
\int\limits_{0}^{2\pi}d\phi \ exp\left \{
\frac{A}{D} \Bigl[ \left( x_b-x_a\right)\ sin(\phi)-Ap\ sin^2(\phi)\Bigr]
\right \}
\end{equation}
Taking into account $A/D >> 1$ we can evaluate the integral by the
steepest descent method and obtain a simple formula
\begin{equation}
\label{eq118} \Delta \approx
\frac{\sqrt D}{2
\sqrt{\pi A\Bigl[\left(x_b-x_a\right)/2-
A p\Bigr]}}
exp\Biggl \{ \frac{A}{D}\Bigl[x_b-x_a -
A p\Bigr]\Biggr \}
\end{equation}
where $p$ is the constant given by (\ref{eq116}).
For the case of CP we have $p\approx 1.632$
while for the QP we have  $p\approx 2.132$. For both of them we have
$x_b-x_a = 1$.
In the considered range $D << A << \sqrt {D}$ the expression under
 the square root in the denominator can not be zero at physically
reasonable values of the noise intensity $D \leq 10^{-1}$. The formulas
(\ref{eq116}) and (\ref{eq118}) are the second result of the present
manuscript.

\section{Conclusions}
The results obtained testify that
the perturbation expansion for the action $\alpha$ in the modulation
amplitude $A$ yields
a reasonable and convergent expression at least in the low frequency limit.
In this case we restrict ourselves by the second order term
$O\left(A/D\right)$. In our opinion the corrections from the third and
higher order contributions (which are $O\left(A^2/D\right)$) will not distort
the results appreciably except in a narrow region near the bifurcation
point $A_c$. In this region our results diverge with the scaling laws
obtained in \cite{Ryv04}, \cite{Dyk04}, \cite{Dyk05}, \cite{Dyk051}. For
the regime of weak modulation our results precisely coincide with those
known from the literature. The formula (\ref{eq118}) in the low limit of
its validity range $D << A << \sqrt {D}$ ($A$ small enough for the terms
 $O(A^2)$ to be neglected) with taking into account that $x_b-x_a=1$ for
both the CP and QP yields the formula (6.34) from \cite{Jun93}
obtained for the case of the QP in the limit of small driving frequencies.
The value $x_b-x_a$ is a coefficient before the linear in $A/D$ term
in the exponent of the escape rate
enhancement and is
actually the so called logarithmic susceptibility from the \cite{Sme99},
\cite{Sme99a}, \cite{Ryv04}, \cite{Dyk04}, \cite{Dyk05}, \cite{Dyk051}
theory. The value $1$ is in agreement with
 $\chi (0)=lim_{\Omega \rightarrow 0}
\left[\pi \Omega/sh\left(\pi \Omega\right)\right]=1$ obtained in \cite{Sme99}
for the case of the CP.

From Fig.1, Fig.2, Fig.3 and Fig.4 we see that the present theory
yields analytical description for the retardation of the exponential
growth of the escape rate enhancement (i.e., transition from a log-linear
regime to more moderate growth). Moreover Fig.2 exhibits the examples of
the reverse behavior when the escape rate enhancement attains a maximum
at some value of $A/D$ and then becomes to decrease with
the further increase of $A/D$. It is worthy to note that this
phenomenon vividly manifests itself for the QP and is not noticed
for the CP. Regretfully the \cite{Leh00},
 \cite{Leh000},
\cite{Leh03} and \cite{Sme99}, \cite{Sme99a}, \cite{Ryv04},
\cite{Dyk04}, \cite{Dyk05}, \cite{Dyk051} theories were not exemplified by
the case of the QP and our prediction can not be directly compared with the
results of those theories.

From the comparison of Fig.1, Fig.2 and Fig.3, Fig.4 respectively we see
that at a given $A/D$ the periodic driving produces stronger
 escape rate enhancement for the case of CP than that of QP.
 The latter certainly can not be
explained by the fact that the barrier height for the CP ($1/6$) is smaller
than that for the QP ($1/4$) because the barrier height does not enter the
formulas (\ref{eq115}) and (\ref{eq118}). This phenomenon can be attributed
to the only difference between these potentials manifested in the value of
the frequency near the bottom of the well $\omega_a$. For the QP this value
($\sqrt 2$)
is higher than that for the CP ($1$). The latter means that the QP goes steeper
from the bottom of the well that hinders the escape rate enhancement
by periodic driving.
Thus the precise shape of the potential is of utmost importance for the
phenomenon of interest.

We obtain two forms of the resulting formula valid for arbitrary potentials with an
activation barrier. The formula (\ref{eq115}) encompasses the
case of arbitrary $A/D$ except a narrow region near the bifurcation
point. Its drawback is that it is very cumbersome.
Nevertheless it can be easily tackled by a computer with the help of,
 e.g., {\it Mathematica}. Its main merit is that it contains only the
notions used at initial setting the problem and can be used by people
engaged in applications of the theory without reading the rest parts of the present
manuscript. The formula (\ref{eq118})
is valid for the case of moderately strong modulation
 $D << A << \sqrt {D}$. Its merit is that it has closed and quite
 simple analytical form to be used by practitioners in chemistry and
 biochemistry for quick by hand estimates. The exponent in this formula
 explicitly and vividly demonstrates how the linear term $O\left(A/D\right)$
dominating at weak modulation (comparatively small $A/D$) and
 responsible for the
log-linear regime is replaced by that substracted by the term
$O\left(A^2/D\right)$ at further increase of $A/D$  providing the
reardation of the exponential growth of the escape rate enhancement.
Regretfully we can not directly
compare our results with those of the \cite{Leh00}, \cite{Leh000},
\cite{Leh03} theory because the latter is inapplicable for the case of slow
modulation.

The formula (\ref{eq118}) is valid in a rather narrow range of the parameters.
However this range turns out to be relevant for applications in enzymology.
The typical
value of noise intensity at enzymatic reactions is
$D\approx 3 \cdot 10^{-3}$ \cite{Sit06}. In this case we have the range of
validity for (\ref{eq118}) as $ 1 << A/D << 18$. The typical values of $A/D$
for a particular model suggested in \cite{Sit06} were estimated as
$A/D \approx 10 $ and find themselves at some stretch within this
 range. The stringent results for this case are given by the formula
(\ref{eq115}) and can be seen in Fig.3 and Fig.4.

We conclude that for the particular case of the low frequency limit the aim to obtain
tractable and convenient formulas describing the escape rate enhancement by
periodic driving seems to be attained. The mathematical tools used at their deriving are
within the scope of elementary methods.

\section{Appendix}
We denote
\begin{equation}
\label{eq119} R =\frac{1}{\Gamma_0(q_b(t))}
\int\limits_{q_b(t)}^{x_c}ds\ exp\left(\frac{U(s)}
{D}\right)s-
\frac{1}{\Gamma_0(x_a)}
\int\limits_{x_a}^{x_c}ds\ exp\left(\frac{U(s)}
{D}\right)s
\end{equation}
then
\begin{equation}
\label{eq120} S(q_b(t))\approx -(q_b(t)-x_a)+ R
\end{equation}
Our aim here to show that the term $R$ is $O\left(\sqrt D \right)$
then (\ref{eq56}) will be proven. Let us write
\begin{equation}
\label{eq121} R=\frac{1}{2\lambda}\frac{\partial}{\partial x_b}
ln\Biggl \{\int\limits_{q_b(t)}^{x_c}ds\ exp\left[-\lambda (s-x_b)^2\right]/
\int\limits_{x_a}^{x_c}ds\ exp\left[-\lambda (s-x_b)^2\right]
\Biggr \}
\end{equation}
Making use of the substitution $r/\sqrt \lambda=s-x_b$ we obtain
\begin{equation}
\label{eq122} R\approx \frac{1}{2\lambda}\frac{\partial}{\partial x_b}
ln\Biggl \{\int\limits_{\sqrt \lambda \left (q_b(t)-x_b\right)}^{\infty}
dr\ exp\left(-r^2\right)/
\int\limits_{-\infty}^{\infty}dr\ exp\left(-r^2\right)
\Biggr \}
\end{equation}
As a result of straightforward calculations we have
\begin{equation}
\label{eq123} R\approx \frac{D}{\omega_b^2 Q_0\left(q_b(t)\right)}
exp\left(-\frac{\omega_b^2 A^2 sin^2(\phi)}
{2D}\right)
\end{equation}
Taking into account (\ref{eq53}) we obtain the required result
\begin{equation}
\label{eq124} R \sim \sqrt D
\end{equation}

Acknowledgements.  The work was supported
by the grant from RFBR.

\newpage

\begin{figure}
\begin{center}
\includegraphics* [width=\textwidth] {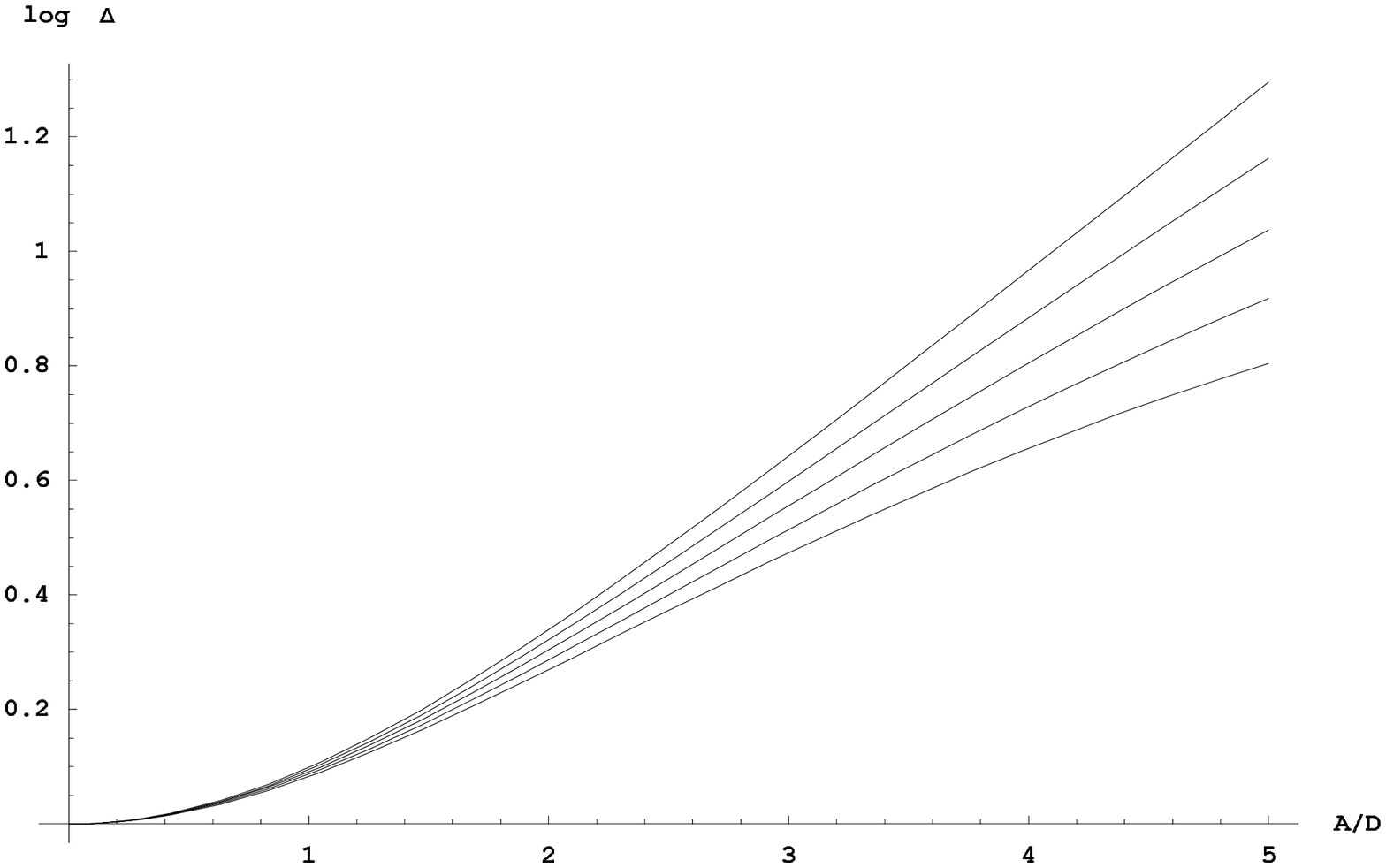}
\end{center}
\caption{The dependence of the escape rate enhancement on
the driving amplitude to noise intensity ratio for the case of
cubic (metastable) potential $U(x)=x^2/2-x^3/3$. The values
of the noise intensity $D$ from the down
line to the upper one respectively are:
  $5\cdot 10^{-2}$; $4\cdot 10^{-2}$;$3\cdot 10^{-2}$;
$2\cdot 10^{-2}$; $1\cdot 10^{-2}$.} \label{Fig.1}
\end{figure}

\clearpage
\begin{figure}
\begin{center}
\includegraphics* [width=\textwidth] {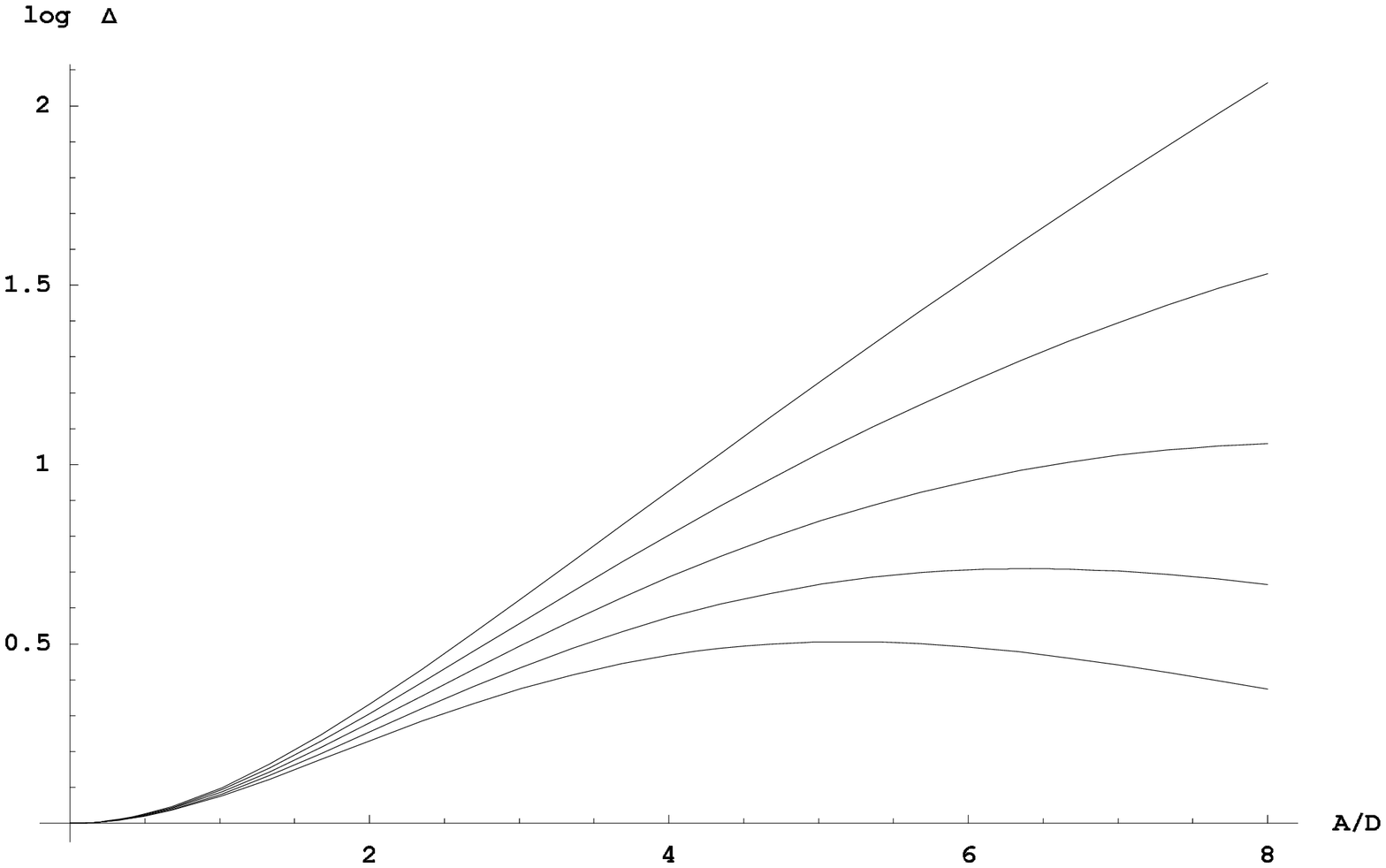}
\end{center}
\caption{The dependence of the escape rate enhancement on
the driving amplitude to noise intensity ratio for the case of
quartic (bistable) potential $U(x)=-x^2/2+x^4/4$. The values
of the noise intensity $D$ from the down
line to the upper one respectively are:
  $5\cdot 10^{-2}$; $4\cdot 10^{-2}$;$3\cdot 10^{-2}$;
$2\cdot 10^{-2}$; $1\cdot 10^{-2}$.}
\label{Fig.2}
\end{figure}

\clearpage
\begin{figure}
\begin{center}
\includegraphics* [width=\textwidth] {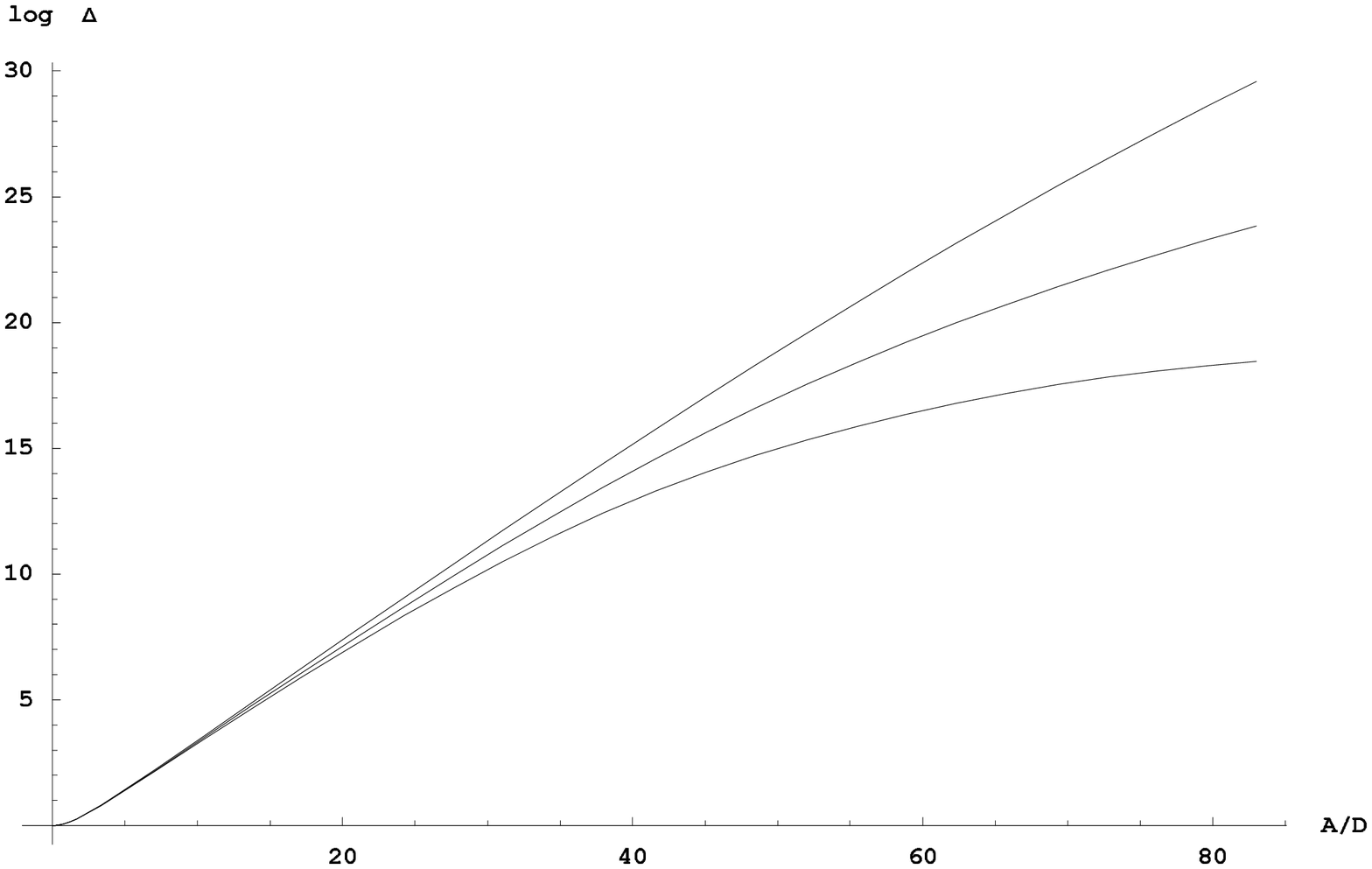}
\end{center}
\caption{The dependence of the escape rate enhancement on
the driving amplitude to noise intensity ratio for the case of
cubic (metastable) potential $U(x)=x^2/2-x^3/3$. The values
of the noise intensity $D$ from the down
line to the upper one respectively are:
  $3\cdot 10^{-3}$; $2\cdot 10^{-3}$;$1\cdot 10^{-3}$.}
\label{Fig.3}
\end{figure}

\clearpage
\begin{figure}
\begin{center}
\includegraphics* [width=\textwidth] {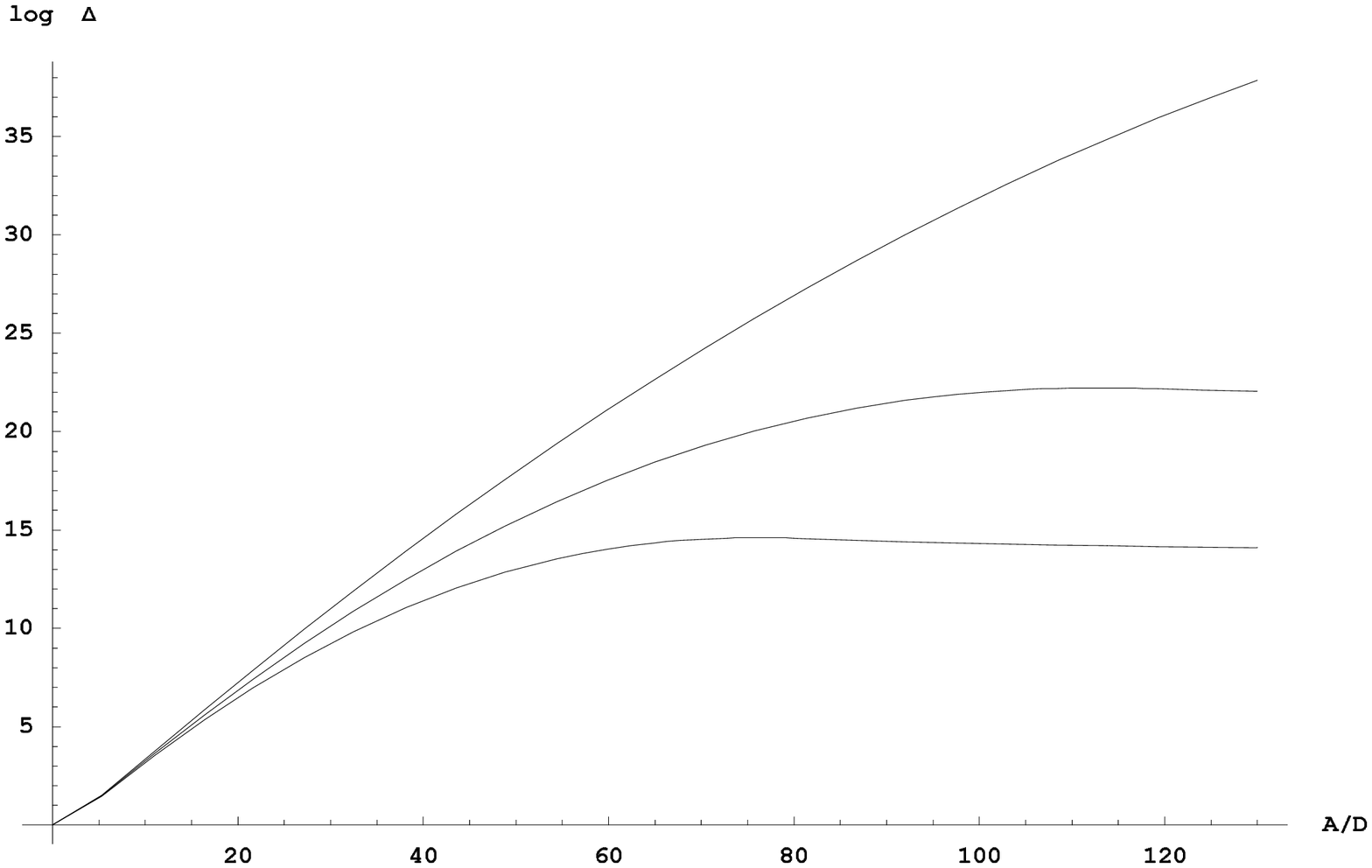}
\end{center}
\caption{The dependence of the escape rate enhancement on
the driving amplitude to noise intensity ratio for the case of
quartic (bistable) potential $U(x)=-x^2/2+x^4/4$. The values
of the noise intensity $D$ from the down
line to the upper one respectively are:
  $3\cdot 10^{-3}$; $2\cdot 10^{-3}$;$1\cdot 10^{-3}$.}
\label{Fig.4}
\end{figure}

\end{document}